\documentclass[acmsmall,screen]{acmart}

\AtBeginDocument{%
  }

\setcopyright{acmlicensed}
\acmJournal{PACMHCI}
\acmYear{2025}
\acmDOI{XXXXXXX.XXXXXXX}

\acmConference[MobileHCI'25]{the ACM on Human-Computer
Interaction}{September 22--26, 2025}{Sharm El-Sheikh, Egypt}
\acmISBN{978-1-4503-XXXX-X/18/06}

\usepackage{enumitem}

\begin{document}

\title{How Problematic are Suspenseful Interactions?}


\author{Alarith Uhde}
\email{uhde@fc.ritsumei.ac.jp}
\orcid{0000-0003-3877-5453}
\affiliation{%
  \institution{Ritsumeikan University}
  \city{Osaka}
  \country{Japan}
}

\renewcommand{\shortauthors}{Uhde et al.}

\begin{abstract}

  Current ``social acceptability'' guidelines for interactive technologies
  advise against certain, seemingly problematic forms of interaction.
  Specifically, ``suspenseful'' interactions, characterized by visible
  manipulations and invisible effects, are generally considered be problematic.
  However, the empirical grounding for this claim is surprisingly weak. To test
  its validity, this paper presents a controlled replication study ($n=281$) of
  the ``suspensefulness effect''. Although it could be statistically replicated
  with two out of three social acceptability measures, effect sizes were small
  ($r\leq.2$), and all compared forms of interaction, including the suspenseful
  one, had high absolute social acceptability scores. Thus, despite the slight
  negative effect, suspenseful interactions seem less problematic in the
  overall scheme of things. We discuss alternative approaches to improve the
  social acceptability of interactive technology, and recommend to more closely
  engage with their specific social situatedness.

\end{abstract}



\received{6 February 2025}
\received[revised]{8 May 2025}
\received[accepted]{29 May 2025}

\maketitle

\section{Introduction}%~1 page

%TODO: "Suspensefulness effect" benennen

% Breiterer Kontext:
% 1. Wenn Menschen 
% 1. Es wäre schön, wenn wir SA Interaktionen einfach nach einer Hand voll
%    eindeutiger Prinzipien gestalten könnten.
% 2. subtlety (aber: Uhde et al., 2022)
%   - suspenseful gestures are not always suspenseful (Uhde et al.

Research about social acceptability in Human-Computer Interaction (HCI) aims to
improve how people experience interactions with technology in social
situations. Broadly speaking, this research develops guidelines to design such
interactions, with the goal to create positive or at least non-negative
experiences for users and witnesses. Given the wide range of technology-based
interactions we find in social situations today, this research and the
resulting guidelines potentially have a far-reaching impact on many peoples'
everyday lives.

A major focus of current social acceptability guidelines are recommendations
about the form of interaction. Among them, one of the most frequently cited
guidelines is the ``suspensefulness effect''. This effect states that
``suspenseful''~\citep{Reeves2005} interactions, characterized by \emph{visible
manipulations} (e.g., pressing a button on a key to unlock a car), and
\emph{invisible effects} (i.e., no ``response'' of the car, such as flashing
headlights), should be avoided (e.g.,~\citep{Montero2010, Ens2015,
Koelle2020}).  Suspenseful forms of interaction have been associated with low
social acceptability from the user's perspective~\citep{Montero2010} and
described as ``awkward''~\citep{Ens2015}.

Although a state-of-the-art guideline (e.g.,~\citep{Koelle2020}), the
suspensefulness effect has a surprisingly weak empirical grounding. It is
mainly based on one study~\citep{Montero2010} with a small sample size
($n=16$).  Retrospectively, we can also assess some problems with the study
design, as outlined below.  Moreover, recent work indicated that at least other
people (i.e., not the user) do not find suspenseful gestures generally
unacceptable~\citep{Uhde2022d, Uhde2023a}.  In other words, we currently work
with a design guideline for social acceptability that has a weak and partially
inconclusive empirical basis.

This paper contributes a validity test for the suspensefulness effect. It
builds on the pioneering work by~\citet{Montero2010}, but controls for various
confounding variables, includes additional social acceptability measures, and
uses a larger sample size ($n=281$) to increase confidence in the findings. So
far, this is the first (and mostly successful) attempt to replicate the
suspensefulness effect. As a second contribution, however, we contextualize the
effect and its practical relevance for interaction design. Although differences
were statistically significant, the effects were small (all $r\leq.2$), and all
tested gestures including the suspenseful variant had high absolute social
acceptability scores across all measures. This indicates that other factors
beyond the mere visibility of manipulations and effects may be more relevant
for how users experience an interaction.  Finally, we discuss alternative
approaches  for practitioners and recommend a closer engagement with the
specific social situations they design for.

\section{Background}

\subsection{Social Acceptability of Interactions with Technology}

%foreshadow things here that I pick up in the discussion: Transparency,
%device-directedness

Social acceptability (also ``social acceptance'') is a widely used, yet
somewhat ill-defined concept. Many publications in the field do not provide an
explicit definition, and among those that do, there is no common agreement. One
position, introduced in an early work by~\citet{Rekimoto2001}, is that ``input
devices should be as natural and (conceptually) unnoticeable as possible for
use in various social settings'' (see also~\citep{Kwok2023}). This definition
implies disturbing effects of the interaction on others who notice it. The
suggested solution is essentially to ``hide'' the interaction. While pragmatic,
this also circumvents the social acceptability problem by removing the
interaction from the social situation, instead of making it more
acceptable~\citep{Uhde2022d}. The definition is also quite narrow, and cannot
explain for example clearly visible, yet commonly accepted interactions with
technology, such as working with a laptop in a café or using a vending machine.
Moreover, in some social situations like face-to-face conversations, ``hidden''
design patterns can create more negative experiences than open and transparent
forms of interaction~\citep{Uhde2022d}.

%TODO: The critique of ‘unclear’ definitions of ‘self-image’ and ‘external
%image’ seems misplaced to me. The example provided afterward effectively
%highlights the social complexities involved in self vs external images, and
%how they can differ and influence the social acceptability depending on the
%context. The authors should consider rephrasing or clarifying this point.

\citet{Koelle2020} provide an overview of further definitions, and later settle
on their own. They highlight that interactions should be ``consistent with the
user's self-image and external image, or alter them in a positive way''. This
position addresses the social more explicitly.  Unlike~\citet{Rekimoto2001},
\citet{Koelle2020} focus more on what the interaction expresses about the user.
The definition remains a bit abstract, however, and introduces further concepts
(e.g., ``self-image'', ``external image'') that are not self-evident.  Think
about a teenager using Virtual Reality (VR) goggles in front of both their
friends and parents, for example. They may all have different, perhaps
conflicting ideas of how the interaction would have to change to create a more
positive image (e.g., more ``audacious'' vs. ``funny'' vs. ``well-behaved'').
In other words, it is not clear what constitutes these ``images'' and how they
could become more positive.  A second shortcoming of this working definition is
that it does not account for the social situatedness of the interaction, which
can have a drastic effect on its social acceptability.  For example, consider
the different experiences that phone calls typically create in libraries,
compared with parks or rock concerts (see e.g.,~\citep{Uhde2021b, Uhde2025a};
also~\citep{Rico2010}).

Nevertheless, \citet{Koelle2020}'s definition makes a critical distinction
between the social acceptability from the user perspective and from the
perspective of potential witnesses (see also~\citep{Alallah2018}). In the case
of suspenseful interactions, for example, \citet{Montero2010} found low social
acceptability from the user perspective, whereas~\citet{Uhde2022d, Uhde2023a}
found positive experiences from the witness perspective.

For the present paper, we use a broad definition of social acceptability,
focusing on the goal to promote positive and avoid negative experiences in
relation to interactions with technology in social situations, while
acknowledging its specific social situatedness.

\subsection{Relationship Between Form of Interaction and Experience}

Given the goal to design interactions that create positive or less negative
experiences, we need to better understand how interactions and experiences
relate to each other. For that purpose, \citet{Reeves2005} have introduced an
influential taxonomy that links characteristics of the form of an interaction
with experiential effects on witnesses. The taxonomy uses basic variations of
(in)visibility of an interaction's ``manipulations'' and ``effects'', resulting
in a 2$\times$2 matrix with four different categories.  We already introduced
``suspenseful'' interactions above, with \emph{visible} manipulations and
\emph{invisible} effects.  Following the same example activity of unlocking
a car, we can alternatively design an interaction as ``magical'', with an
invisible manipulation (e.g., pressing the car key inside a bag, hidden from
witnesses) and visible effects (e.g., the headlights flash). The two remaining
variants are ``expressive'' interactions, with visible manipulation and
effects, and ``secretive'' interactions, where both manipulations and effects
are invisible.  \citet{Reeves2005} suggested this taxonomy as a design space to
create certain experiences for spectators of staged performances, for example
at a rock concert. However, this taxonomy was later adopted for interactions
around witnesses more broadly.

\subsection{Unacceptable Suspenseful Gestures}

%TODO: Better argue why suspensefulness can also be useful

The claim that suspenseful gestures have a lower social acceptability than
\citet{Reeves2005}'s other three categories is primarily based on one lab study
by \citet{Montero2010}. The authors designed a total of eight phone-based
interactions, two for each category. They then presented videos of these
gestures to sixteen participants. Finally, they asked participants an open
question to assess the gestures' social acceptability from the witness
perspective, and two closed questions to assess the social acceptability from
the user perspective.  The central finding is based on an analysis of one of
the closed questions (about ``public spaces'', see below). All in all, they
found that both suspenseful gestures from their set had a lower social
acceptability score than the two expressive, secretive, and magical gestures.

This study has been impactful in the social acceptability literature, and
directly led to further design inspirations. For example, \citet{Ens2015} built
on this finding with their concept of ``candid'' interactions, which redesign
suspenseful into expressive forms by making effects more visible.
\citet{Rico2010} suggested the other path away from suspenseful and towards
secretive interactions, by making the manipulation less visible and more
``subtle''.  Moreover, \citet{Ahlstrom2014} explored if suspenseful
interactions that remain suspenseful might still be acceptable with the
appropriate size and duration of the manipulation, and they concluded that the
manipulation should be small and not last for long.

\citet{Montero2010}'s study also served as a post-hoc explanation for an
earlier finding (e.g., in~\citep{Koelle2020}): witnesses who overheard phone
calls, where they could only hear one side of the conversation
(``halfalogues''~\citep{Emberson2010}), experienced them as more disturbing
than ``full conversations'' between two people~\citep{Monk2004a}. However,
there are competing explanations for this ``halfalogue'' effect. One is that
phone calls go along with increased loudness, compared to face-to-face
speech~\citep{Forma2012}. Another explanation is that halfalogues imply
a ``need-to-listen'' for the witnesses~\citep{Monk2004b, Norman2014}.
Finally,~\citet{Emberson2010} argue that halfalogues are more disturbing,
because they are less predictable than full dialogues or monologues. Crucially,
their finding that halfalogues, but not monologues, had disturbing effects on
witnesses, is inconsistent with predictions based on the suspensefulness effect
alone.

Despite this impact on new design approaches and potential reinterpretation of
earlier research, follow-up studies to confirm the suspensefulness effect
itself are still missing. In addition, some recent work indicated that
suspenseful interactions can also have positive effects, at least on the
witness experience in certain social situations~\citep{Uhde2022d, Uhde2023a}.
In other words, to clarify the validity of the suspensefulness effect and its
role for the social acceptability of interactions, a replication study seems
warranted.

From today's perspective, we can identify a few issues with the original study
design. First, the interactions used in the study do not only differ in their
visibility patterns. For example, one of the expressive gestures is ``Throwing
money [into a] vending machine'', which may be more familiar to people than the
suspenseful gesture ``Writing big letters in the air''. Thus, the activity
represented by the different gestures and their familiarity are potential
confounding variables.  Second, as pointed out recently~\citep{Uhde2023a},
there is a conceptual issue with how~\citet{Reeves2005}'s taxonomy defines
``visibility'', which also affects \citet{Montero2010}'s study. We can define
visibility based on visual perception alone, by asking ``can the witness see
the manipulation/effect?''.  However, with this definition, the ``secretive''
category does not make sense: If the witness can neither see the manipulation
nor the effect, they would not see the interaction at all. Thus, they would not
experience it at all, instead of having a ``secretive'' experience.
Alternatively, we can define visibility as ``standing out from the
conventional''. A manipulation may be defined as ``visible'' if it draws
attention, for example because it is unusual. But this comes with other
problems.  Think of a conventional phone call (``halfalogue''), which would
then fall into the ``secretive'' category---neither the user's voice nor the
lack of a reply as such draw attention for being ``unusual''. Conversely,
\emph{muting} the user's voice, for example with a ``Silent Speech
Device''~\citep{Li2019}, would render the interaction ``suspenseful''
(see~\citet{Uhde2023a} for a more detailed argument of this issue). For the
rest of this paper, we will use the ``perception-based'', not the
``convention-based'' definition. Based on this, we argue that the ``secretive''
gestures from~\citet{Montero2010} (the videos are available
online\footnote{\url{https://dl.acm.org/doi/suppl/10.1145/1851600.1851647/suppl_file/p275-montero.mp4}}),
is effectively suspenseful: the manipulations are clearly visible. But the fact
that they have high social acceptability scores, similar to the magical and
expressive gestures, in a sense negates the supposed suspensefulness effect.
Finally, the small sample size reduces statistical confidence in the findings.
Small samples are not uncommon in the HCI literature, especially given that
\citet{Montero2010}'s work was a pilot study.  However, the community has not
produced validation studies for the effect so far, despite its criticality for
social acceptability guidelines.

% Also: ANOVA statt KW

In the ensuing development in the literature, some further issues arose. First,
there are now various social acceptability measures in use, and it is not clear
whether \citet{Montero2010}'s findings hold with these other measures. Second,
as \citet{Uhde2022d} noted, many everyday interactions with smartphones are in
fact ``suspenseful'', which indicates that real world users find at least some
suspenseful interactions socially acceptable. And finally, excluding
suspenseful interactions altogether from the design repertoire significantly
reduces the design space, and currently it is not clear whether that is
justified. Strictly speaking, every interaction where we cannot fully hide the
manipulation from witnesses, with smartwatches, phones, or laptops for example,
would also need to show them some effect to make them socially acceptable. This
seems impractical, and conflicts with some definitions of social acceptability
itself~\citep{Rekimoto2001}.

In sum, although \citet{Montero2010}'s study provided a valuable contribution
by first making the connection between an interaction's visibility pattern and
its social acceptability, fifteen years later we still cannot conclusively say
whether this claim actually holds true. Yet, it has been widely adopted in the
literature and found its way into current social acceptability guidelines.

\section{Replication Study}

The main goal of this study was to replicate the suspensefulness
effect~\citep{Montero2010}, which states that suspenseful gestures are less
socially acceptable than the other forms from \citet{Reeves2005}'s taxonomy
(i.e., expressive, magical, and secretive).  It was approved by our institute's
ethics committee and
preregistered\footnote{\url{https://aspredicted.org/65N_Q76}}.  All data
(anonymized), material, and analysis scripts are available in the supplementary
material\footnote{See ACM Digital Library or
\url{https://doi.org/10.5281/zenodo.15571897}}. We tested the following
hypotheses:

\begin{description}
  \item[H1:] Suspenseful interactions are less socially acceptable than all
   three other forms, using the same measure as \citet{Montero2010}.
  \item[H2:] Suspenseful interactions are less socially acceptable than all
three other forms, when measured with two other common social acceptability
    scales:
    \begin{enumerate}[label=\textbf{\alph*)}]
      \item The social acceptability scale by \citet{Pearson2015}
      \item The social acceptability scale by \citet{Koelle2018}
    \end{enumerate}
\end{description}

\noindent{}In addition, we included further items for exploratory analysis as
described below.

\subsection{Method}

\subsubsection{Participants}

%TODO: Report demographics in more detail

We used G*Power~\citep{Faul2007} to estimate the required sample size given
a medium effect ($f=.25$), $\alpha=.05$ and $\text{a priori test power}=.95$,
using the one-way ANOVA setting\footnote{For the actual analysis, we opted for
non-parametric testing because all scales are based on single items.}. This
indicated a necessary sample size of 280, and accordingly, we recruited 280
participants via prolific.com. We screened participants for English language
fluency and selected a sample based in the United Kingdom, similar to the
original study. The study had an estimated length of 5 minutes, with
a compensation of 0.9~GBP (i.e., 10.80~GBP per hour). One additional
participant was marked as ``timed-out'' on prolific, but later completed the
study.  Following the preregistered procedure, we also included them in our
analysis.  Thus, our final sample size was 281 (106 male, 174 female,
1 diverse/other; $age_{mean}=37.8$, $age_{sd}=12.0$).

\begin{figure*}
  \includegraphics[width=0.98\linewidth]{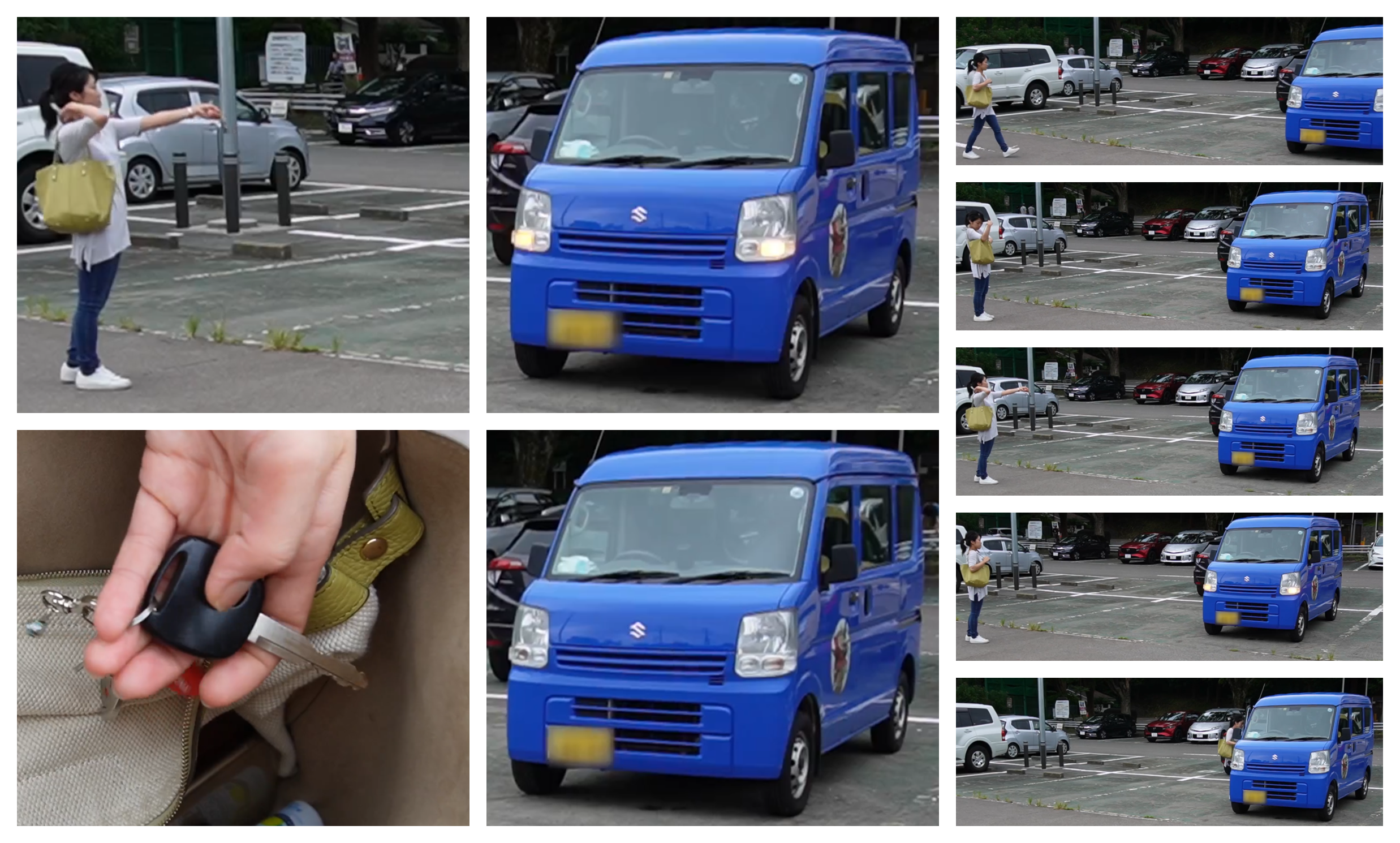}
  \Description{A combined figure of nine photographs that show a person
  unlocking a car.}
  \caption{Key elements of the interaction used in the study. The four larger
  pictures illustrate the interaction. On the left, they show the visible
  (top) and invisible manipulation (bottom), from the witness perspective.
  The two middle pictures illustrate the visible effect (top) and invisible
  effect (bottom). The right side illustrates an example flow of the scene from
  top to bottom. First, the person approaches the car. Second, she checks her
  bag and in this case takes out the keys. Third, she points the keys at the
  car to unlock it. Fourth, the car ``reacts'' with flashing headlights. Fifth,
  she enters the car. The full videos can be found in the supplementary
  material.}%
\label{fig:car}
\end{figure*}

\subsubsection{Procedure}

%TODO: How did the authors ensure participants perceived the gesture as
%intended?

Participants accessed our study through a link to limesurvey.net. They first
read an introduction and gave their informed consent about data analysis and
participation. On the next page, they read a short description of the scenario.
They would see a person approaching their car and using a remote key to unlock
it. We then randomly assigned them to one of four conditions to watch a short
video (15-17s) of that interaction, with an expressive, magical, secretive, or
suspenseful gesture (see Figure~\ref{fig:car}). We used the same base
interaction for all gestures to avoid confounding effects, and only varied the
visibility of the manipulation and effect. In the two ``visible manipulation''
conditions (suspenseful and expressive), the person openly pointed the key at
the car, with an extended arm. In the ``invisible manipulation'' conditions
(magical and secretive), she pressed the key inside her purse (thus, invisible
from the witness perspective). The visible effect were flashing front lights of
the car (in the expressive and magical conditions). In the invisible effect
conditions (suspenseful and secretive) the front lights did not flash.

Following this video, we included a comprehension check question to verify if
participants perceived the gesture as intended, with the (in)visible
manipulations and effects from a witness perspective that matched their
condition. Fourty-seven participants ($17\%$) answered the comprehension check
incorrectly, which exceeded our preregistered threshold of $10\%$. Thus, we
reran the analysis to check whether the results would change if we a)
reassigned participants to the visibility pattern they declared in their
comprehension checks (instead of their actual groups) or b) excluded the
participants who failed the checks from the analysis. Both tests did not affect
the statistical significances of the hypothesis testing, and both only had an
effect on the pairwise comparisons of the exploratory tests of the
``confidence'' scale. We have included a footnote below with the details, and
share the data, full procedure, and detailed results in the supplementary
material. After the comprehension checks, participants filled in the four
different social acceptability measures described below, a question each about
their gender and age, and finished the study.

\subsubsection{Measures}

To date, there is no single established scale for social acceptability, and
many studies rely on self-developed items~\citep{Koelle2020}. However, a few
scales from the literature have been used multiple times before and thus allow
for some empirical comparison. We referred to the suggested scales from
\citet{Koelle2020}'s literature survey, but excluded three scales that seemed
unsuitable for our study, because they specifically focus on wearables and
people with disabilities~\citep{Profita2016, Kelly2016, Kim2006}. Including
\citet{Montero2010}'s scale, this left us with four social acceptability
measures.

\paragraph{Social Acceptability Scale from Montero et al. (2010)}

\citet{Montero2010} gave participants two items to measure the social
acceptability from the user's perspective: ``How would you feel performing this
gesture in the following situations? a. in public places b. at home''. They
measured these with a 6-point scale ranging from 1 (``Embarrassed'') to
6 (``Comfortable''). Similar to their own study, we only included scale a) for
hypothesis testing and used scale b) for exploratory analysis.

\paragraph{Social Acceptability Scale from Pearson et al. (2015)}

Second, we included \citet{Pearson2015}'s single item: ``To what extent do you
feel this interaction is socially acceptable?'', with a 5-point scale from
``completely unacceptable'' to ``completely acceptable''.

\paragraph{Social Acceptability Scale from Koelle et al. (2018)}

Third, we included two items from \citet{Koelle2018}: ``How acceptable would it
be to perform the presented gesture in public?'' (social acceptability) and
``How comfortable would you feel performing this gesture in an everyday public
setting, such as a busy sidewalk?'' (confidence). Responses are measured on
a 7-point Kunin scale~\citep{Kunin1955}. We only included the social
acceptability scale for hypothesis testing and used the confidence scale for
the exploratory analysis.

\paragraph{Audience-and-Location Axes (Rico and Brewster, 2010)}

Finally, we included the ``Audience-and-Location Axes'' (ALA;~\citep{Rico2010})
as an exploratory measure. The ALA include six locations (home, pavement or
sidewalk, driving, passenger on a bus or train, pub or restaurant, workplace)
and six audiences (alone, partner, friends, colleagues, strangers, family).
Participants rate how they would feel performing a gesture at these locations
and in front of those audiences. We made a few changes to the scale. First, we
included three further locations we considered relevant for the specific
interaction of unlocking a car: ``on an open, public car park'', ``in
a multi-storey car park'', and ``at a motorway service area''. Second, unlike
the original version which used a binary measure, we used a 5-point scale
ranging from 1 ``very socially uncomfortable'' to 5 ``very socially
comfortable'' (similar to~\citep{Alallah2018}). Third, we included an
additional response option in case users find the location or audience
unsuitable for the interaction: ``this location/audience does not make sense to
me for this interaction''.

\subsection{Results}

% Assumption: The suspenseful gesture was not less socially acceptable than the
% other gestures

\subsubsection{H1 confirmed: suspenseful gesture least acceptable}

\begin{figure*}
  \includegraphics[width=0.98\linewidth]{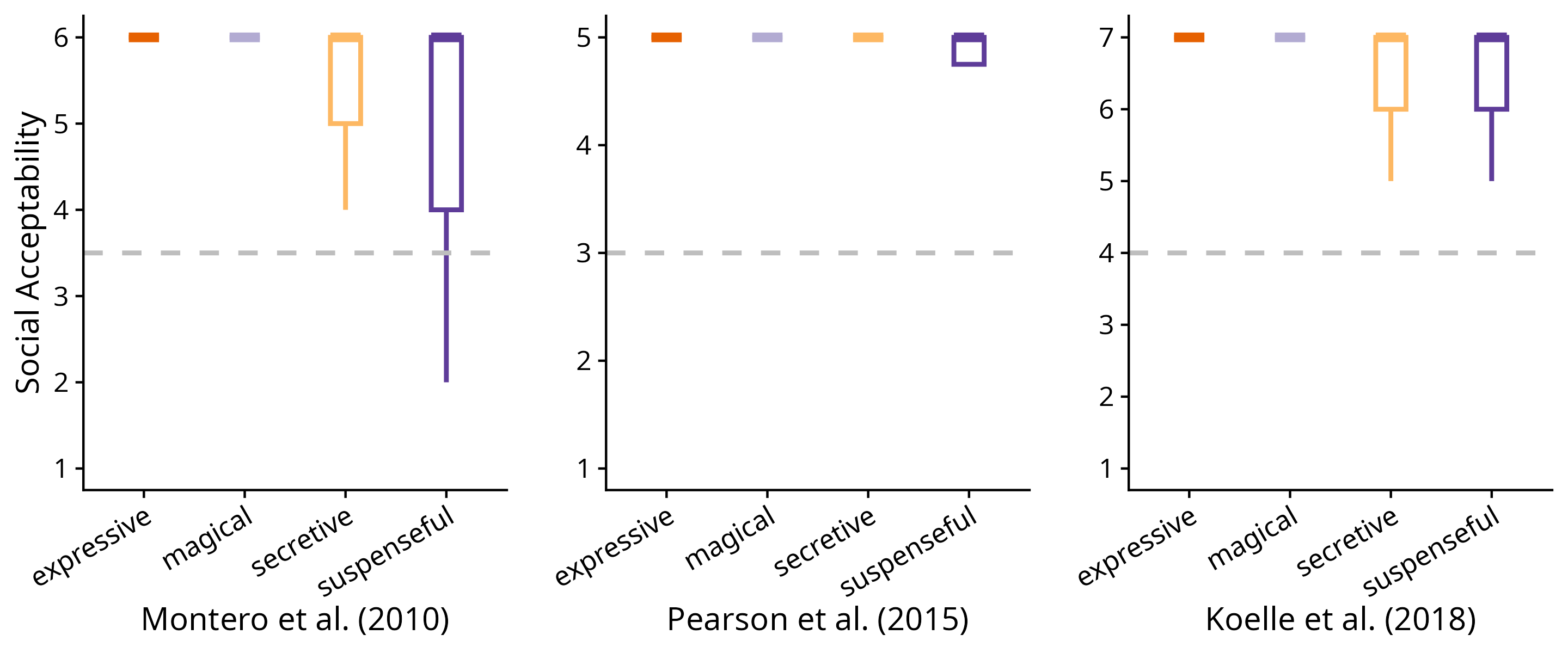}
    \Description{Three boxplots, each depicting the scores of the four
    interaction for one of the three social acceptability measures. The boxes
    of the expressive and magical interactions are thin lines at the upper end
    of each scale. The box for the secretive interaction is slightly larger on
    the scales by Montero et al. and Koelle et al., but the whisker still ends
    above the center of the scales. Finally, the box of the suspenseful gesture
    is the largest across all three plots. Nevertheless, the interquartile
    range remains above the centers of the scales, and the whisker only cuts
    the center in the scale by Montero et al.}
  \caption{Boxplots of the four interactions' scores on the three social
  acceptability measures. The dashed line represents the center of the
  respective scale.}%
\label{fig:sa}
\end{figure*}

A Kruskal-Wallis-Test with gesture type as independent variable (expressive,
magical, secretive, suspenseful) and \citet{Montero2010}'s social acceptability
measure as dependent variable revealed a significant overall effect ($H=15.86$,
$p<.001$, $df=3$; see Figure~\ref{fig:sa}). Thus, we conducted three pairwise
comparisons using Mann-Whitney-U-Tests, which confirmed that participants
experienced the suspenseful gesture as less socially acceptable than the
expressive ($p<.01$, $r=.18$), magical ($p<.01$, $r=.19$), and secretive
($p<.05$, $r=.13$) gestures. In sum, this confirms hypothesis H1: using
\citet{Montero2010}'s original measure, the suspenseful gesture was less
socially acceptable than the other three gestures.

\subsubsection{H2 a) not confirmed but similar direction: suspenseful gesture less acceptable than expressive, magical}

A second Kruskal-Wallis-Test with gesture type as independent variable and
\citet{Pearson2015}'s social acceptability measure as dependent variable also
revealed a significant overall effect ($H=11.00$, $p<.05$, $df=3$). Again,
three pairwise comparisons using Mann-Whitney-U-Tests confirmed that
participants experienced the suspenseful gesture as less socially acceptable,
compared with the expressive gesture ($p<.01$, $r=.17$) and the magical gesture
($p<.05$, $r=.12$). However, the difference with the secretive gesture was not
significant ($p=.13$). Thus, H2 a) could only partially be confirmed: Using
\citet{Pearson2015}'s measure, the suspenseful gesture was not less socially
acceptable than all three other gestures (specifically the secretive gesture).

\subsubsection{H2 b) confirmed: suspenseful gesture least acceptable}

In a third Kruskal-Wallis-Test with gesture type as independent variable, we
used \citet{Koelle2018}'s social acceptability measure as dependent variable,
which again revealed a significant overall effect ($H=16.64$, $p<.001$,
$df=3$). Pairwise comparisons using Mann-Whitney-U-Tests confirmed that
participants experienced the suspenseful gesture as less socially acceptable,
compared with the expressive ($p<.01$, $r=.16$), magical ($p<.001$, $r=.20$),
and secretive ($p<.05$, $r=.13$) gestures. This confirms hypothesis H2 b):
Using \citet{Koelle2018}'s measure, the suspenseful gesture was less socially
acceptable than all other three gestures.

\subsubsection{Hypothesis testing summary: Suspensefulness effect mostly confirmed}

Despite the concerns about the empirical basis of the suspensefulness effect
indicated above, our study confirms a negative effect of suspenseful gestures,
compared with expressive, magical, and secretive gestures, from the user's
perspective in two out of three tests. The third test also confirmed that the
suspenseful gesture was perceived as less acceptable than the expressive and
magical gestures. Although the difference to the secretive gesture was not
significant, we draw the overall conclusion from this study that suspenseful
gestures can lead to (slightly) lower social acceptability scores, from the
user's perspective. We adopt this interpretation for the rest of the paper, in
part because it contrasts our a priori skepticism. However, we will come back
to this in the discussion.

\subsubsection{Exploratory analysis}
%further items

\paragraph{Suspenseful gesture still highly acceptable}

The above analyses may seem to imply that suspenseful gestures are socially
unacceptable and should be avoided. But when looking at absolute scores, we get
a different picture (see Figure~\ref{fig:sa}). The scores, including those of
the suspenseful interaction, were overall positive, on all scales. To confirm
this, we ran Median-based, robust one-sample tests (the \emph{onesampb}
procedure from~\citep{Wilcox2012}), with a bootstrap sample size of
($n=10000$), to check whether the scores were significantly above the center of
the three scales. All three differences were significant
($\text{Median}_{Montero}=6$, $\text{Scale Center}_{Montero}=3.5$,
$p_{Montero}<.01$; $\text{Median}_{Pearson}=5$, $\text{Scale
Center}_{Pearson}=3$, $p_{Pearson}<.01$; $\text{Median}_{Koelle}=7$,
$\text{Scale Center}_{Koelle}=4$, $p_{Koelle}<.01$). Thus, despite relative
differences among the gestures, even the ``least acceptable'', suspenseful
gesture was overall socially acceptable.

\paragraph{ALA: suspenseful gesture socially acceptable in typical locations, audiences}

In a further analysis using the ALA data for the suspenseful gesture, we first
checked which locations and audiences participants found unsuitable for the car
unlocking interaction from our study. Our criterion was that at least half of
the participants would consider a location or audience unsuitable
($\text{critical }n=34$). This was the case for the locations ``driving''
($n=53$), ``passenger on a bus or train'' ($n=51$) and ``pub or restaurant''
($n=36$). This seems plausible, as unlocking a car may not be a meaningful
activity when already inside a moving vehicle or while in a pub or
restaurant---independent of the possible social acceptability of such a gesture
in these situations. Therefore, we excluded these locations from further
analysis. Participants found all other locations ($\text{max }n=13$) and
audiences ($\text{max }n=3$) suitable for this interaction.

\begin{figure*}
  \includegraphics[width=0.8\linewidth]{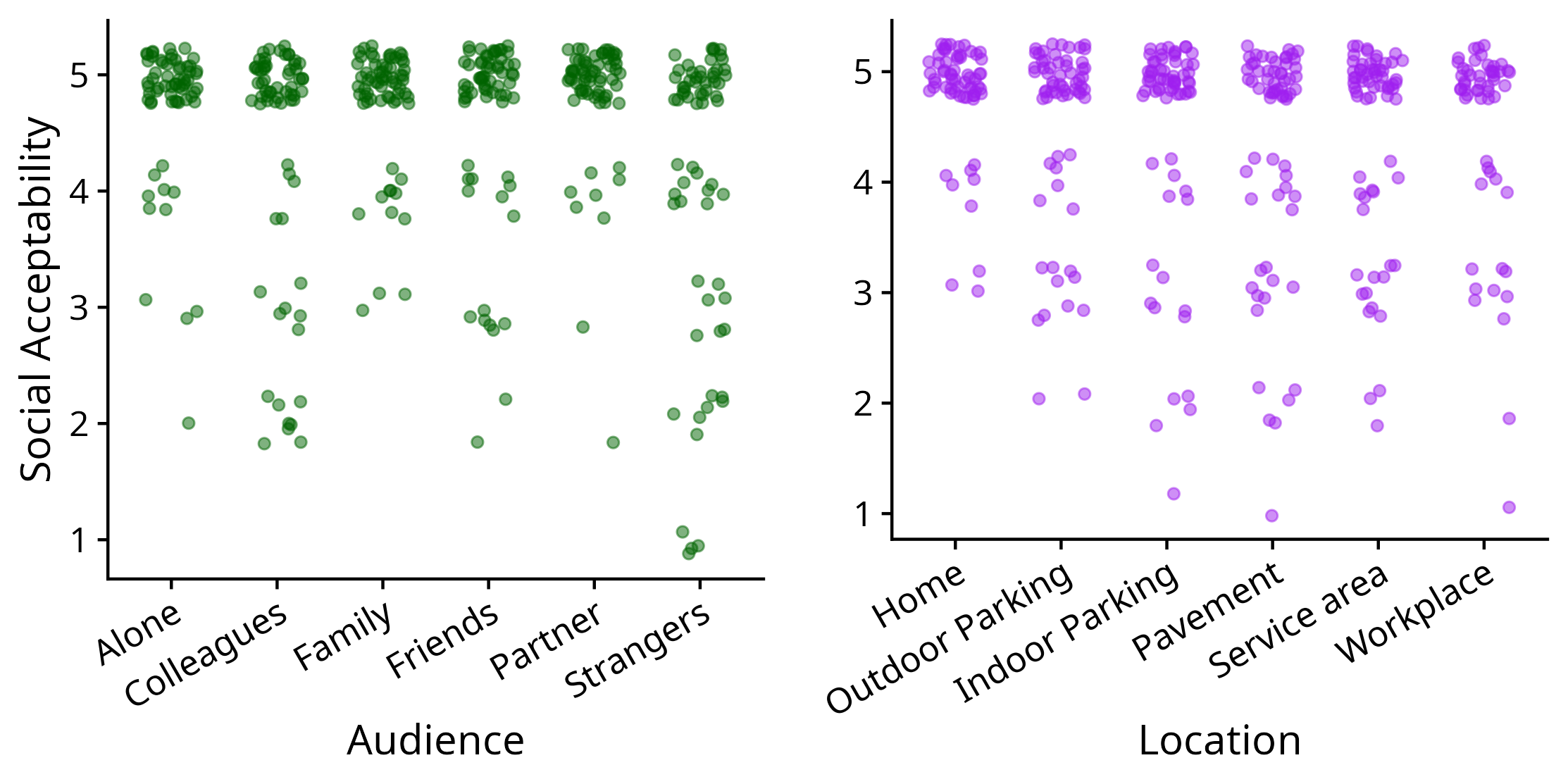}
  \Description{Two scatter plots of participants' ratings of the suspenseful
  gesture on the Audience-and-Location Axes. For each audience and location,
  there is a dot cloud at the maximum of each scale (5), with fewer dots
  scattered on the lower scores. Most scores are greater or equal to 3, the
  center of the scale. There are also a few dots at the lowest end of the
  scale, for ``strangers'' (4), and ``indoor parking'', ``pavement'', and
  ``workplace'' (1 each).}
  \caption{Scatter plot of the suspenseful gesture's social acceptability score
  for the six suitable locations and all audiences. Note that all scores are
  based on one item only and thus natural numbers; some horizontal and vertical
  jitter was added to reduce overlap.}%
\label{fig:ala}
\end{figure*}

Across all remaining locations and audiences, median social acceptability was
at the scale maximum ($Median=5$). Thus, we decided to only conduct a basic
visual analysis. Figure~\ref{fig:ala} shows scatter plots of all scores for the
suspenseful interaction and illustrates the high levels of social acceptability
across locations and audiences. Most scores are at the upper end of the scale,
and only few scores reached below the center ($3$) of the scale, indicating
generally high social acceptability.

\paragraph{Additional Scales: No difference ``at home'', small effect on
``confidence''}

Finally, we analyzed the remaining two items. A Kruskal-Wallis-Test with
gesture type as independent variable and \citet{Montero2010}'s ``at home'' item
as dependent variable revealed no significant effect ($H=5.10$, $p=.16$,
$df=3$). Although not specified precisely, participants may have understood
``home'' as a suitable location, as indicated by the ALA analysis
above---perhaps in the sense of a private parking lot, a garage, or in front of
their house.

A second Kruskal-Wallis-Test with gesture type as independent variable and
\citet{Koelle2018}'s confidence item revealed a significant main effect
($H=17.00$, $p<.001$, $df=3$). Pairwise comparisons with family-wise error
correction using the Hochberg method~\citep{Hochberg1988} revealed that
participants gave lower confidence ratings for the suspenseful gesture compared
with the magical ($p<.001$, $r=.20$) and secretive ($p<.05$, $r=.10$) gestures,
but not the expressive ($p=.11$) gesture\footnote{This is where the results
from the comprehension check analyses (reassigning/excluding participants with
incorrect responses) differed slightly. The pairwise comparison between the
suspenseful and magical interaction was not significant ($p=.06$ in both
cases). Additionally, in the ``reassigning'' analysis, the pairwise difference
between the suspenseful and expressive interaction was significant ($p<.05$,
$r=.12$). For details, see supplementary material.}.

\section{Discussion}

This paper reported a replication study of the ``suspensefulness
effect''~\citep{Montero2010}, which states that users experience
``suspenseful'' interactions as less socially acceptable than ``expressive'',
``magical'', and ``secretive'' interactions. Building on the pioneering work
by~\citet{Montero2010}, our study controlled for several issues with the
initial experiment that left alternative explanations open. First, we used
a larger sample size ($n=281$ vs. $n=16$) to increase the overall confidence in
the findings.  Second, we held the activity (unlocking a car) constant across
all conditions and only varied the form of interaction, to eliminate the
alternative explanation that the effect is based on differences between
activities, not their form. Third, we designed the ``secretive'' interaction
with an invisible manipulation (from the witness perspective), further reducing
ambiguity. In the initial experiment, the ``secretive'' interactions could also
be categorized as suspenseful, but they had high social acceptability scores.
And finally, we used a standardized online study with prerecorded videos and
random assignments of participants to groups, which eliminates further
potential confounding factors such as systematic group differences or
experimenter effects. With these checks in place, the suspensefulness effect
could still be (mostly) confirmed, with small but significant main effects on
all three measures and eight out of nine significant pairwise differences in
line with the assumptions. The only exception was a non-significant difference
between the suspenseful and secretive interaction on~\citet{Pearson2015}'s
social acceptability scale, which we discuss further below. In sum, this study
validates the suspensefulness effect and eliminates various sources of
skepticism.

That said, our analysis of the absolute social acceptability scores puts the
importance of the effect into perspective. Despite small differences between
the different forms of interaction, all of them had high social acceptability
scores from the user perspective, close to the upper end of all scales.
Moreover, as the ALA analysis shows, median social acceptability was at the
upper end of the scale for all audiences and locations deemed reasonable. In
sum, users found the suspenseful interaction slightly less, but still highly
acceptable, particularly in locations where it makes sense to perform it.

In addition, note that the effect reported in this paper only covers how
potential users experience the interaction. Earlier studies had already
reported suspenseful interactions with overall positive experiences from the
perspective of witnesses, in certain social situations such as face-to-face
conversations~\citep{Uhde2022d, Uhde2023a}. In addition, we can see in common,
everyday interactions, that several suspenseful interactions such as taking
a photo or text messaging with a smartphone seem generally acceptable for many
people.

Taken together, we think that for all practical purposes, the current general
recommendation in the literature to avoid suspenseful forms of interaction is
not timely anymore. Some suspenseful interactions are certainly ``awkward'', as
reported earlier~\citep{Montero2010, Ens2015}. But this awkwardness is not
inherent to suspenseful interactions. As the high overall acceptability scores
from our study show, suspenseful interactions can be perfectly acceptable. To
design interactions suitable for various social situations, we need alternative
approaches as discussed further below.

\subsection{Remaining Inconsistencies}

A few statistical effects were not fully in line with our assumptions.  Most
notably, the difference between the suspenseful and the secretive gesture on
\citet{Pearson2015}'s social acceptability scale was not significant.  One
reason may be that~\citet{Pearson2015} uses only a 5-point scale, whereas
\citet{Montero2010} and \citet{Koelle2018} use 6-point and 7-point scales,
respectively. This numerical difference may have left a possible small existing
difference undetected.  Another reason may be the wording of this scale, which
ranges from ``completely unacceptable'' to ``completely acceptable''.
Acceptability is a slightly more technical concept, compared with the
``embarrassed/comfortable'' scale from~\citep{Montero2010} and the (pictorial)
Kunin scale from~\citep{Koelle2018}. In addition, the scale requires
participants to make a judgement about the more general ``acceptability'' of
the interaction, rather than their own subjective experience, which may be more
challenging.

A second inconsistency is that the difference in ``confidence'' between the
suspenseful and the expressive interaction was not significant.  One reason
could be that these two interactions both include the spatious hand gesture,
which might draw attention on the user. However, the follow-up analyses using
reassignments and exclusions of participants based on the comprehension checks
led to slightly different results. In sum, this finding remains inconclusive
and may need further investigation in the future.

\subsection{Limitations}

%TODO Check, maybe add:
%- online study, not real life
%- implicit effect (opening the car door; problem also in previous work
%  reference)
%- particular scenario as reference

This study also has some limitations. First, one problem with this and previous
social acceptability studies is the lack of standardized measures. We included
\citet{Montero2010}'s original measure and two relatively established,
additional measures~\citep{Pearson2015, Koelle2018}. However, all three
measures only consist of a single item each. In addition, they have not been
constructed based on a theoretical model of social acceptability. This has to
do with the current unclarity about what exactly we mean when we talk about
social acceptability (see~\citep{Uhde2022d, Uhde2025a} for more detailed
discussions of this).

%TODO: Oben noch das neue IWC-Paper rein

A second problem is that the tested interaction only represents a single
scenario. We do not claim that the high overall scores generalize to all other
social situations.  In fact, our point is quite the opposite: Our findings show
that the relevance of the suspensefulness effect does not generalize, and we
provide evidence for that. Suspenseful interactions can be highly acceptable
from the user perspective. Our ALA analysis indicates some transferability of
that effect to other locations and audiences, with our specific interaction.
But we would like to highlight that the takeaway should not be that everyone
should now use suspenseful interactions. Instead, we need to think beyond the
form of interaction as an isolated factor that determines social acceptability.
Further replication studies with other suspenseful interactions that vary for
example in ``awkwardness'' or ``familiarity'' in a certain situation are still
needed for a better understanding of how the form relates to social
acceptability.

Third, we designed this study as an online experiment to be able to standardize
the experimental conditions and specifically only vary the form of interaction,
with everything else kept constant. We made this choice to minimize potential
confounding effects, which was one of the central issues
with~\citet{Montero2010}'s study. The downside of this approach is that
participants did not perform the interactions themselves, which threatens
ecological validity. In their literature survey, \citet{Koelle2020} noted
a need for more social acceptability studies performed in more natural
settings, which we did not address here.

Finally, we only compared the four ``main'' categories from
\citet{Reeves2005}'s taxonomy (suspenseful, expressive, magical, secretive),
and omitted the more nuanced variants. For example, users can also ``partially
reveal'' a manipulation/effect, or amplify it. We made this choice to keep the
findings comparable with~\citet{Montero2010}. But future work could provide
further insights into these more nuanced forms of interaction and their effects
on situated experiences.

\subsection{Design Implications: Revisiting Recommendations Against Suspenseful
Interactions}

Previous work has already indicated that suspenseful interactions may not be as
problematic as claimed in current guidelines. Besides the widely accepted
smartphone interactions and the positive witness experiences already mentioned,
the first to make that point were in fact \citet{Reeves2005} with their
taxonomy. They provided examples for suspenseful forms of interaction that
create ``suspense'' in a positive way, such as the tension before a roller
coaster ride. Arguably, \citet{Montero2010} provided further evidence for this,
with their ``secretive'' interactions that can be seen as in fact suspenseful,
and which participants perceived as socially acceptable. The present study adds
empirical evidence, controlling for these issues with the experimental design.

But even if the effect seems negligible---is it a problem to continue advising
against suspenseful interactions? We would argue that it is. To illustrate this
point, consider typical interactions with smartphones. Current suspenseful
interactions strike a balance between pragmatic design and privacy.  Experience
has shown that hiding the fact somebody is touching their smartphone display
does not seem necessary in most situations. But redesigning it in an
unobservable way would be challenging.  Conversely, hiding or at least not
broadcasting the ``effects'' (what happens on the display) allows for a certain
degree of privacy. In other words, here we have an entirely acceptable,
suspenseful form of interaction, which appropriately solves a specific problem
in various social situations, but the literature advises against it.  Of
course, we could also say that smartphone interactions have found their way
into practice anyway, despite current guidelines, so they did not hurt either.
But if we want such design guidelines to have any meaningful impact on design
practice, we need to develop them along with empirical evidence and not guide
designers in the wrong direction.

A perhaps less obvious but similarly problematic and related aspect is that
social acceptability guidelines are currently dominated by a focus on the form
of interaction.  This distracts from other, arguably more important factors.
Mobile phone calls serve as a good illustration for this problem. Even if we
only look at the ``same'' form of interaction in different situations, people
experience phone calls quite differently~\citep{Uhde2021b}. They can be very
disturbing when they hinder other people from doing whatever they are up
to---be it reading, or ``making the caller's hair'' (see~\citep{Uhde2025a} for
further examples).  Conversely, the same phone call may be fun for others to
listen in to if they are bored, or serve as a ticket to talk~\citep{Sacks1992,
Olsson2020} that leads to a new friendships~\citep{Uhde2025a}. The form is not
the main issue here, and it has different effects, depending on situational
needs. Thus, instead of tweaking visibility or audibility patters, we think
that designers need to engage more closely with actual target situations, and
study how the interaction fits in.  To illustrate this point, the noise
produced by a phone call (i.e., ``visible/audible manipulation'') can be
a problem in a library, so redesigning it to be less noisy (e.g., with a Silent
Speech Device~\citep{Li2019}) could reduce negative experiences.  Users already
do that by opting for alternative, text-based communication.  However, this
design change is unlikely to help in other social situations, like face-to-face
conversations. Here, the noise is not the core problem that leads to negative
experiences, but the phone call itself signals a lack of interest in the
conversation partner~\citep{Uhde2025a}. In other words, there is no general
``form'' that is suitable across all social situations, and purely form-based
recommendations need to be taken with a grain of salt.

Other factors beyond the mere visibility of manipulations and effects may
contribute more strongly to how users and witnesses experience an interaction.
This includes the novelty of the interaction, user and witness personalities,
their mood and situational roles, how the interaction fits in with other
situated practices, local and legal regulations, and many more
(see~\citet{Uhde2025a} for an overview).  Thus, although non-suspenseful
interactions can be slightly more acceptable, we recommend that practitioners
prioritize these other factors during interaction design and engage more
closely with the target situations.

\section{Conclusion}

This study reported a replication of \citet{Montero2010}'s ``suspensefulness
effect'', which states that users experience suspenseful interactions as less
socially acceptable than other forms. The effect was replicated, but the
findings also show that it may be of less practical relevance than it is
currently treated. All interactions, including the suspenseful interaction, had
high absolute social acceptability scores. Based on this and further findings
from other studies, the recommendation against suspenseful interactions still
found in current social acceptability guidelines~\citep{Koelle2020} should be
dropped. Instead, researchers and designers should engage more closely with
actual situations of use and look for ways to improve the ``fit'' between their
interaction and specific social situations.

\begin{acks}
This study was funded by Tokyo College, the University of Tokyo.
\end{acks}
%TODO
% Check order of suspenseful, expressive, secretive, magical. Alphabetical?

\bibliographystyle{ACM-Reference-Format}
\bibliography{/home/alarith/Documents/bibliography}%

%\appendix



\end{document}